# A New Equation for Activity Calculation in Pulse Irradiation: Derivation, Simulation and Experimental Validation

Zaijing Sun

*Nuclear Engineering Program, South Carolina State University, 300 College Street NE, Orangeburg SC 29117*

**ABSTRACT**

To calculate the radioactivity of product nuclides generated in pulse irradiation, it is generally assumed that the irradiation is approximately continues in the entire irradiation period ($t_i$) and the flux of incoming irradiation particle can be obtained by averaging their intensity in each pulse period ($T$). However, this approximation fails to acknowledge the fact that the product nuclides are not created in each pulse period ($T$) evenly: they are only produced in a very short pulse width ($t_p$) and then decay in a relative long rest time ($t_r = T - t_p$). Given by the enormous number of pulses, the sum of these decays may not be negligible. To make the activity calculation in accordance with the real situation in pulse irradiation, we scrutinize the details of irradiation and decay processes in each pulse, applies the geometric series to obtain the activity superimposition of millions of pulses, and derives a novel activity equation particularly suitable for pulse irradiation. The experimental results, numerical simulations, and activity measurements from photon activation driven by a pulsed electron LINAC have confirmed the validity of this new equation. The comparison between the new and traditional equations indicates that their discrepancy could be significant under certain conditions. The limitations of the new activity equation for pulse irradiation are discussed as well.

**KEYWORDS:** activity equation; geometric series; continue irradiation; pulse irradiation.

## INTRODUCTION

Radioactivity calculation is a primarily concern in nuclear activation analysis, medical isotope production, health physics, and other fields in nuclear and radiological sciences. Historically, the activity may be generated from sources which are capable of continuous radiation, such as nuclear reactors and radioisotope resources. The number of product nuclides $N(t_i)$ at end of irradiation time $t_i$ follows the equation

$$N(t_i) = \frac{R}{\lambda} \cdot \left(1 - e^{-\lambda t_i}\right) = \frac{N_0 \varphi \sigma}{\lambda} \cdot \left(1 - e^{-\lambda t_i}\right) \tag{1}$$

where $R$ is the reaction rate of selected nuclear reaction, $N_0$ is the original number of target nuclides, $\lambda$ is the decay constant of product nuclides, $\varphi$ is the particle flux of the irradiation particle (e.g. high energy photon or thermal neutron), and $\sigma$ is the cross section of corresponding nuclear reaction. The product $\varphi\sigma$, or reaction rate density, is usually an integral over the energy of the incoming particle from threshold energy which the reaction occurs to the maximum energy ($\varphi\sigma = \int_{E_{threshold}}^{E_{max}} \varphi(E)\sigma(E)dE$). The burn-up of target nuclides is usually very small and could be ignored in most situations. The total number of radionuclides $N$ follows the curve in Figure 1a, which reaches its peak $N(t_i)$ at the end of irradiation and then decays according to the exponential law. The activity at this moment always equals the peak number of product nuclides times the decay constant

$$A(t_i) \equiv \lambda N(t_i) = R\left(1 - e^{-\lambda t_i}\right) = N_0 \varphi \sigma \left(1 - e^{-\lambda t_i}\right) \tag{2}$$

Nowadays, due to the development of accelerator technology, more and more radioisotopes are produced by pulse irradiation [1]. Traditionally, to calculate the activity from pulse irradiation, scientists usually expand the usage of continuous irradiation Equations (1) and (2) into applications of pulse irradiation without too much consideration of details in pulses. Instead of peak flux in the pulse width, they use average particle flux during the whole irradiation period to calculate the final radioactivity [2-5]. This practice does make sense at the first thought: the repetition rate of pulses is usually relatively high, the irradiation can be regarded as continuous from the point of whole irradiation period, and the half-life of the product nuclides of interest are usually long even comparable to the whole irradiation time.

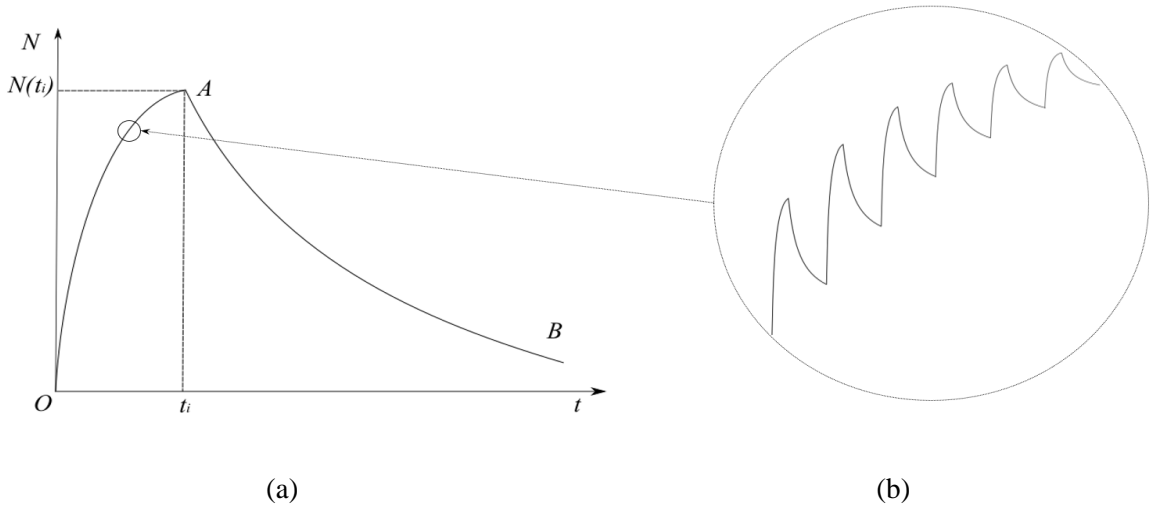

(a) (b)

**Figure 1.** (a) Number of the product nuclides in continuous irradiation assumption; (b) Number of the product nuclides in pulsed irradiation (highly exaggerated).

However, the continuous assumption is questionable when more details of each pulse are revealed. Figure 1a shows the number of product nuclides for a continuous irradiation source such as a nuclear reactor. The product nuclides start to grow from zero when the irradiation starts ($O \rightarrow A$), reach the peak point (maybe close to saturation) at point $A$ when the irradiation stops at $t_i$, and then decay exponentially after irradiation ($A \rightarrow B$). If the irradiation is driven by a pulsed irradiation source (e.g. LINAC), the curve of activation period ($OA$) is not as smooth as in Figure 1a. Imagining we use a magnified glass to see the details of $OA$, it is more like Figure 1b. The zigzag shape (or a series of superimposed bumps) shows growth and decay within the pulses. Even though the overall trend of number of product nuclides is growing, there are drops in pulses, which come from the decay in the rest (or down) time among pulses. Some product radioisotopes have very short half-lives, and the total irradiation period is relatively long. With millions of pulses, the total sum of rest time is not negligible. For instance, in 10 hours irradiation of a typical L-band electron LINAC which has pulse width in microsecond (see Figure 2a), the total up time (addition of millions of pulse-widths) is only in the scale of minutes, and the remaining time (or total rest time) is about 9.9 hours simply for the accelerator to prepare for the next pulse. Given the enormous number of pulses, it looks suspicious to adapt continuous assumption and ignore the decay process during pulses.

What is the activity equation looks like if we adapt the pulsed irradiation model instead? Does the pulse activation model impact significantly on the results of the final activity of radioisotopes? Which physical parameter(s) in pulse irradiation will be significantly affect the final activity of product nuclides? To answer these questions and make the activity calculation more accurate, one needs to establish a new pulse irradiation model in accordance with the real experimental situation in pulse irradiation.

**THEORETICAL: MATHEMATICAL DERIVATIONS OF GENERAL ACTIVITY EQUATION IN PULSE IRRADATION**

In real situation of irradiation, the pulse profile is very complicated (see Figure 2a). It may be the overlap of several sinusoidal waves. In the following discussion, a simplified pulse profile as a rectangular wave in Figure 2b is introduced: $I_p$ is the pulse current, $\bar{I}$ is the average current, $t_p$ is pulse width, $t_r$ is the rest time in each pulse, and $T\ (= t_p + t_r)$ is the pulse period, which is the reciprocal of the repetition rate (or frequency) $f$.

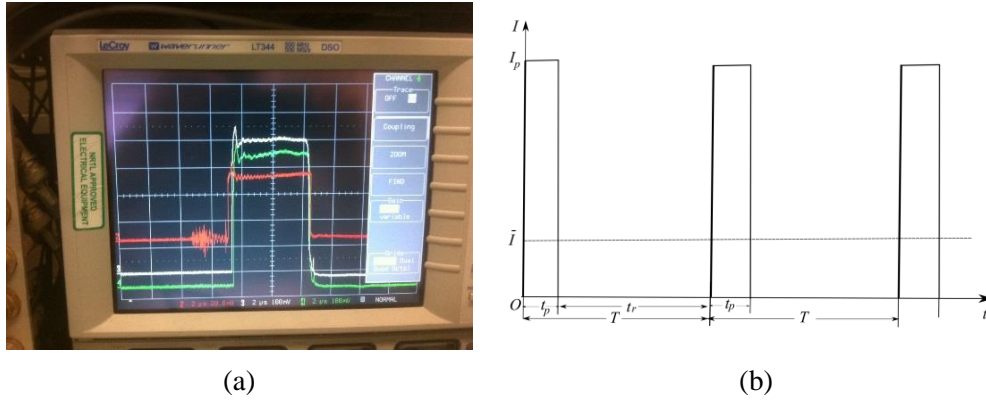

(a) (b)

**Figure 2.** (a) An oscilloscope shows the pulse profile for the LINAC at the Low Energy Accelerator Facility (LEAF) at Argonne National Laboratory; (b) Rectangular wave model for pulse irradiation.

*Derivation 1: Superimposition of activities from each pulse at the end of irradiation*

Initially, it looks like a tedious work to calculate the total number of product nuclides after an irradiation period $t_i$ due to millions of pulses. However, this can be solved by the superimposition of millions of pulses (bumps) using convergent geometry series. The final number of product nuclides is the superimposition of a series of bumps which are illustrated in Figure 2b or in Figure 3. Each bump corresponds to one pulse irradiation. The total number of product nuclides at the moment $t_i$ is the sum of all the residual product nuclides at the moment $t_i$ produced by each pulse. Each pulse creates radioisotopes independently and they are not contact with each other.

For the 1st pulse, it created $N_p$ product nuclides in pulse width time $t_p$. At the moment of $t_i$, product nuclides generated from the first pulse have decayed with a time period of $t_i - t_p$. Thus, according to the decay law, one gets the number of residual product nuclides produced by the 1st pulse at moment of $t_i$ is

$$N_1 = N_p e^{-\lambda(t_i - t_p)} = N_p e^{\lambda t_p} e^{-\lambda t_i} \tag{3}$$

For the 2nd pulse, it created $N_p$ product nuclides in pulse width time $t_p$ as well (the burnup of target nuclide is ignored). At the moment of $t_i$, the decay time for the second pulse is $t_i - t_p - T$. The number of residual product nuclides produced in the 2nd pulse is:

$$N_2 = N_p e^{-\lambda(t_i - t_p - T)} = N_p e^{\lambda t_p} e^{-\lambda t_i} e^{\lambda T} \tag{4}$$

For the 3rd pulse, at the moment of $t_i$, the decay time is $t_i - t_p - 2T$, and

$$N_3(t_i) = N_p e^{-\lambda(t_i - t_p - 2T)} = N_p e^{\lambda t_p} e^{-\lambda t_i} e^{2\lambda T} \tag{5}$$

Accordingly, for the last pulse $m$ before the end of irradiation, the number of residual product nuclides is

$$N_m(t_i) = N_p e^{-\lambda(t_i - t_p - (m-1)T)} = N_p e^{\lambda t_p} e^{-\lambda t_i} e^{(m-1)\lambda T} \tag{6}$$

The total of residual product nuclides at end of irradiation is

$$N(t_i) = N_1 + N_2 + N_3 + \cdots + N_m = \sum_{j=1}^{m} N_j$$

$$= N_p e^{\lambda t_p} e^{-\lambda t_i} + N_p e^{\lambda t_p} e^{-\lambda t_i} e^{\lambda T} + N_p e^{\lambda t_p} e^{-\lambda t_i} e^{2\lambda T} + \cdots + N_p e^{\lambda t_p} e^{-\lambda t_i} e^{(m-1)\lambda T}$$

$$= N_p e^{\lambda t_p} e^{-\lambda t_i}\left(1 + e^{\lambda T} + e^{2\lambda T} + \cdots + e^{(m-1)\lambda T}\right) = N_p e^{\lambda t_p} e^{-\lambda t_i} \sum_{j=1}^{m} e^{(j-1)\lambda T} \quad (7)$$

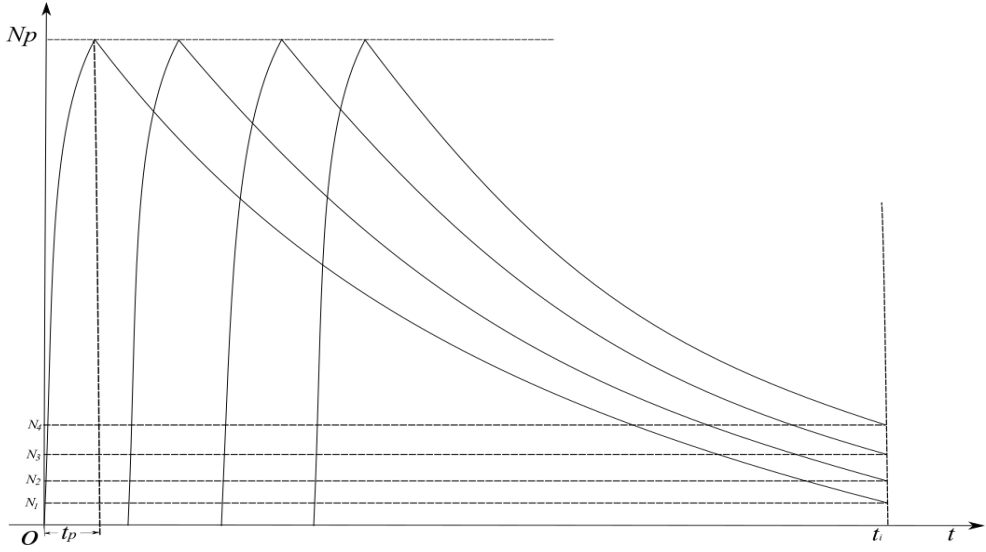

**Figure 3.** Superimposition of pulses (bumps) using geometry series.

Let

$$s = \sum_{j=1}^{m} e^{(j-1)\lambda T} = 1 + e^{\lambda T} + e^{2\lambda T} + \cdots + e^{(m-1)\lambda T} \quad (8)$$

To get the limit of the convergent geometric series above, we multiply $e^{\lambda T}$ on both sides

$$e^{\lambda T} s = e^{\lambda T} + e^{2\lambda T} + \cdots + e^{(m-1)\lambda T} + e^{m\lambda T} = 1 + e^{\lambda T} + e^{2\lambda T} + \cdots + e^{(m-1)\lambda T} + e^{m\lambda T} - 1$$

$$= s + e^{m\lambda T} - 1 \quad (9)$$

By rearranging (9), one gets

$$s = \frac{e^{m\lambda T} - 1}{e^{\lambda T} - 1} \quad (10)$$

The relation of $m$ and $t_i$ is

$$m = int\left(\frac{t_i}{T}\right) \quad (11)$$

By inserting (11) into (10), one obtains

$$s = \frac{e^{int\left(\frac{t_i}{T}\right)\lambda T} - 1}{e^{\lambda T} - 1} \approx \frac{e^{\frac{t_i}{T} \times \lambda T} - 1}{e^{\lambda T} - 1} = \frac{e^{\lambda t_i} - 1}{e^{\lambda T} - 1} \quad (12)$$

Combing (12) with (7), and it yields

$$N(t_i) = \frac{N_p e^{\lambda t_p} e^{-\lambda t_i}(e^{\lambda t_i}-1)}{e^{\lambda T}-1} = \frac{N_p e^{\lambda t_p}(1-e^{-\lambda t_i})}{e^{\lambda T}-1} \tag{13}$$

In the pulse width $t_p$, the irradiation is continuous and $\varphi = \varphi_p$. Combining (13) with (1), one obtains:

$$N(t_i) = \frac{N_0 \varphi_p \sigma e^{\lambda t_p}(1-e^{-\lambda t_p})(1-e^{-\lambda t_i})}{\lambda(e^{\lambda T}-1)} \tag{14}$$

### Derivation 2: Addition of nuclides generated by each pulse

Derivation 1 is based on the superimposition of residual radioactive nuclides produced by each pulse at the end of moment $t_i$. To some extent, it is slightly counterintuitive. The logical way is to follow the time sequences: the growth of activity should be calculated one pulse after another pulse from the start to the end of irradiation. In this section, we will follow this logic and prove that two derivations reach the same result. For the convenience of discussion, we assume that the burn-up of target nuclides is negligible and the number of product nuclides suddenly increased $N_p$ at the end of each $t_p$ (see the sketch in Figure 4).

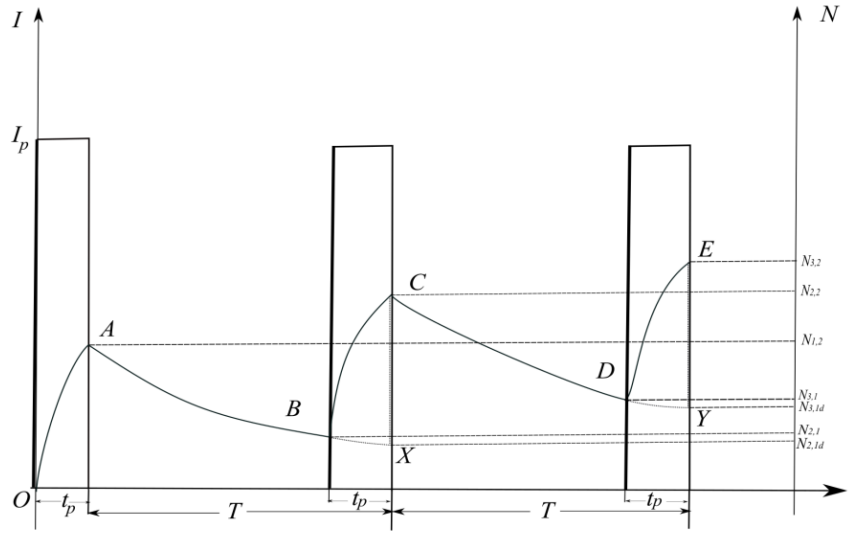

**Figure 4.** Addition of activity generated by each pulse.

For the 1st pulse, $OA$ is a continuous irradiation, at point $O$

$$N_{1,1} = 0 \tag{15}$$

At point $A$, the number of product nuclides is

$$N_{1,2} = N_p \tag{16}$$

For the 2nd pulse, it starts at point $B$

$$N_{2,1} = N_{1,2} e^{-\lambda(T-t_p)} = N_p e^{-\lambda(T-t_p)} = N_p e^{\lambda t_p} e^{-\lambda T} \tag{17}$$

At point $C$, compared with the point $X$ where $B$ decays after a period of $t_p$, the number of nuclides increases $N_p$, thus

$$N_{2,2} = N_{2,1d} + N_p = N_{2,1} e^{-\lambda t_p} + N_p = N_p e^{\lambda t_p} e^{-\lambda T} e^{-\lambda t_p} + N_p = N_p(e^{-\lambda T} + 1) \tag{18}$$

Accordingly, for the 3$^{rd}$ pulse ($DE$)

$$N_{3,1} = N_{2,2}e^{-\lambda(T-t_p)} = N_p(e^{-\lambda T} + 1)e^{-\lambda(T-t_p)} = N_p e^{\lambda t_p}(e^{-2\lambda T} + e^{-\lambda T}) \tag{19}$$

$$N_{3,2} = N_{3,1d} + N_p = N_{3,1}e^{-\lambda t_p} + N_p = N_p(e^{-2\lambda T} + e^{-\lambda T} + 1) \tag{20}$$

For the last pulse $m$ before the end of irradiation, we have

$$N_{m+1,1} = N_p e^{\lambda t_p}(e^{-m\lambda T} + \cdots + e^{-\lambda T}) = N_p e^{\lambda t_p} \sum_{j=1}^{m} e^{-j\lambda T} \tag{21}$$

Similar to derivation 1, we can get the limit of the convergent geometric series as

$$\sum_{j=1}^{m} e^{-j\lambda T} = \frac{1-e^{-\lambda t_i}}{e^{\lambda T}-1} \tag{22}$$

By combining (22), (21) and (1)

$$N_{m+1,1} = \frac{N_0 \varphi_p \sigma e^{\lambda t_p}\left(1-e^{-\lambda t_p}\right)\left(1-e^{-\lambda t_i}\right)}{\lambda(e^{\lambda T}-1)} \tag{23}$$

which is exactly the same as the result of (14).

*General activity equation for pulse irradiation*

Since $T = t_p + t_r$, (14) or (23) can be modified as

$$N(t_i) = \frac{N_0 \varphi_p \sigma e^{\lambda t_p}\left(1-e^{-\lambda t_p}\right)\left(1-e^{-\lambda t_i}\right)}{\lambda(e^{\lambda T}-1)} = \frac{N_0 \varphi_p \sigma\left(1-e^{-\lambda t_p}\right)\left(1-e^{-\lambda t_i}\right)}{\lambda(1-e^{-\lambda T})} \cdot \frac{e^{\lambda t_p}}{e^{\lambda T}}$$

$$= \frac{N_0 \varphi_p \sigma(1-e^{-\lambda t_p})\left(1-e^{-\lambda t_i}\right)}{\lambda(1-e^{-\lambda T})} \cdot e^{-\lambda t_r} \tag{24}$$

And the final activity at the end of pulse is

$$A(t_i) = \frac{N_0 \varphi_p \sigma e^{\lambda t_p}\left(1-e^{-\lambda t_p}\right)\left(1-e^{-\lambda t_i}\right)}{e^{\lambda T}-1} = \frac{N_0 \varphi_p \sigma(1-e^{-\lambda t_p})\left(1-e^{-\lambda t_i}\right)}{(1-e^{-\lambda T})} \cdot e^{-\lambda t_r} \tag{25}$$

If one considers the activity right at the end of last pulse width $t_p$, the decay factor $e^{-\lambda t_r}$ vanishes. This term can be seen as a factor generated by the rest time in the last pulse. Thus, at the end of last pulse width, the total number of product nuclides $N_{total}$ and the corresponding activity $A_{total}$ are

$$N_{total} = \frac{N_0 \varphi_p \sigma\left(1-e^{-\lambda t_p}\right)(1-e^{-\lambda t_i})}{\lambda(1-e^{-\lambda T})} \tag{26}$$

$$A_{total} = \frac{N_0 \varphi_p \sigma(1-e^{-\lambda t_p})\left(1-e^{-\lambda t_i}\right)}{1-e^{-\lambda T}} \tag{27}$$

Equation (27) is the general activity equation for pulse irradiation. If $t_p = T$, or in other word, the irradiation is continuous, then (27) changes its form back to (2), which is the general activity equation for continuous irradiation.

From Equation (27), one can notice that the activity of product nuclides is not only related to $\lambda$, $N_0$, $\varphi$, $\sigma$, but also determined by three time-parameters: $t_p$, $t_i$, and $T$. The final activity in pulse irradiation is proportional to the saturation factor $1 - e^{-\lambda t_p}$ of irradiation in the pulse width $t_p$ and the saturation factor $1 - e^{-\lambda t_i}$ of irradiation in the whole irrational period $t_i$. In addition, it is also inverse proportional to the saturation factor $1 - e^{-\lambda T}$ of irradiation in the pulse period $T$.

**EXPERIMENTAL: PHOTON ACTIVATON WITH PULSED ELECTRON LINAC**

*Photon activation experiments*

To validate the new activity equation in pulse irradiation, photon activation experiments were conducted by the 44 MeV short pulsed LINAC at the Idaho Accelerator Center. Figure 5a is the sketch of the experimental set up, which includes electron gun, the electron-photon converter, and the irradiation sample. Electrons were initially created by hot cathode and then accelerated by a series of alternating RF electric fields in the acceleration cells. Optimized energy around 30 MeV was applied and the total output power was around 2kW. The pulse width is 2.3 µs, the repetition rate is 120 Hz, and the peak current is about 240 mA. The electron beam is focused by magnetic fields to a radius of about 3 millimeters.

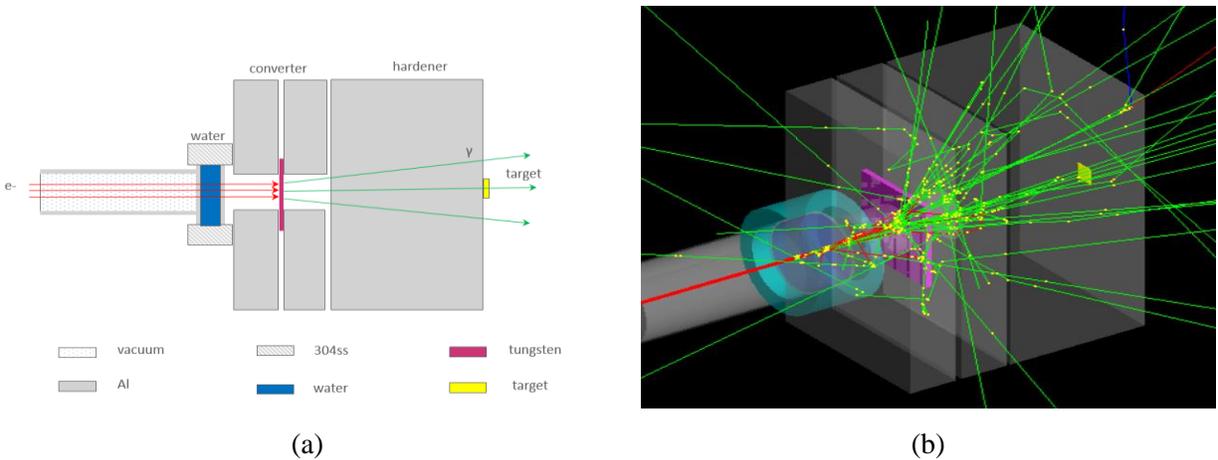

(a) (b)

**Figure 5.** (a) Experimental setup of photon activation driven by a pulsed electron LINAC; (b) Geant 4 simulation of photon shower (view point $\theta = 45°$, $\phi = 135°$).

After the tungsten converter of 3mm thickness, the electron beam completely converted into a bremsstrahlung photon beam. The converter was cooled with forced air continuously to avoid the risk of melting down. The photon flux produced directly after the converter at 30 MeV and 2 kW is approximately $1.1 \times 10^{12}$ photons/sec/cm²/kW. An aluminum hardener of 7.62 cm was positioned after tungsten block to absorb the residual electrons and ensured that the photon beam behind the hardener was predominantly made of high energy photons. A well-known certified reference material, standard reference material 1648a (urban particulate matter) from NIST was served as irradiation target [6] and wrapped with aluminum foil into a 1cm×1cm×1mm square shape. Target was positioned downstream behind the hardener along the beam axis to generate activities for measurements. The total irradiation was lasted 7 hours.

*Gamma ray measurements and spectrum analysis*

After photon activation, the target was cooled down in the accelerator hall for 24 hours to meet the requirement of radiation safety for transferring to the spectroscopy room. Spectra collection was finished by a P-type coaxial detector with 48% efficiency and a resolution less than 1.5keV. After one week, spectra of long-lived isotopes were collected by the same HPGe detector again. Samples were measured in two positions: J and A. Position J is 10cm away from the detector head with an intrinsic peak efficiency of 0.00258@393.529keV for short-lived isotope measurements. Position A is right against the detector with an intrinsic peak efficiency of 0.0625@393.529keV for long lived isotope measurements. Figure 6a indicates a typical gamma spectrum collected by MCDWIN program with some characteristic energy lines and their corresponding radioisotopes [7]. After spectra acquisition, all the gamma spectra files in mp format were input to Gamma-W software for automatic peak analysis (Figure 6b) [8].

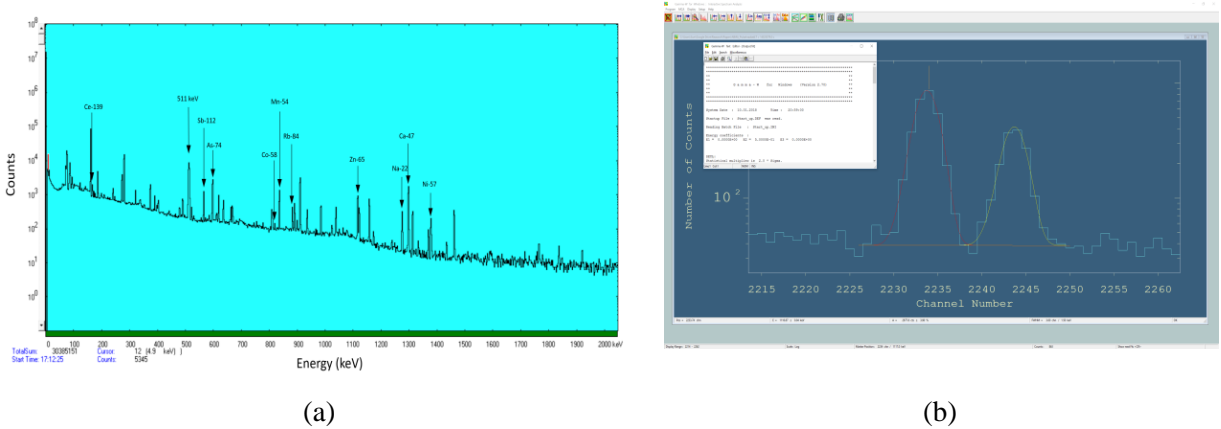

(a) (b)

**Figure 6.** (a) A gamma spectrum collected by MCDWIN (x axis: energy, y axis: counts); (b) Peak analysis with Gamma-W program.

## COMPUTATIONAL: GEANT4 AND MATLAB SIMULATIONS

### Photon flux simulation with GEANT4

To get the photon flux $\varphi$ in the sample, Monte Carlo simulations were performed with GEANT4 toolkit 4.10.3 installed on an HP ProDesk 600 G1SFF workstation running a 64 bits Ubuntu 16.04.1 LTS operating system [9-11]. The simulations followed the exact geometry shown in Figure 5a and generated photon shower illustrated in Figure 5b. The relationship of the track color and its corresponding particle is: photon: green, electron: red, positron: blue, neutron: yellow. The magenta block is the Tungsten converter and the yellow flat cuboid right against the hardener simulates the target.

In the physics list file of the photon shower program, all the electromagnetic processes were added, including G4ComptonScattering.hh, G4GammaConversion.hh, G4PhotoElectricEffect.hh, G4eMultipleScattering.hh, G4eIonisation.hh, G4eBremsstrahlung.hh, G4eplusAnnihilation.hh, and G4ionIonisation.hh. To create an electron beam with measured energy distribution, the default PrimaryGeneratorAction.hh file in the include directory of the program was modified with the class of general particle source (GPS) [12]. A file named "beam.in" stored all the user defined parameters in energy distribution of the beam. Target material was designed as vacuum on purpose to record all the photons entering the cuboid. The output photon.txt recorded all the photons the target can "see" with the information of their position $(x, y, z)$, energy $(E)$, and momentum $(p_x, p_y, p_z)$.

Figure 7a is the energy distribution of the photons entering the target shown in ROOT framework [13]. One can see that the energy distribution of the photons behaves as a typical bremsstrahlung curve: it starts from zero and ends up with the cut-off energy of the incoming electrons. Figure 7b indicates that the photons are dominantly second-generation particles (parentID = 1). Since the first-generation particles are incoming electrons (parentID=0), those photons (parentID = 1) should be created directly by Bremsstrahlung process. Some other generations of photons are also created, but their amount is quite limited compared with that of Bremsstrahlung photons. These photons might be created by other physics processes, such as pair production, photonuclear reactions, etc.

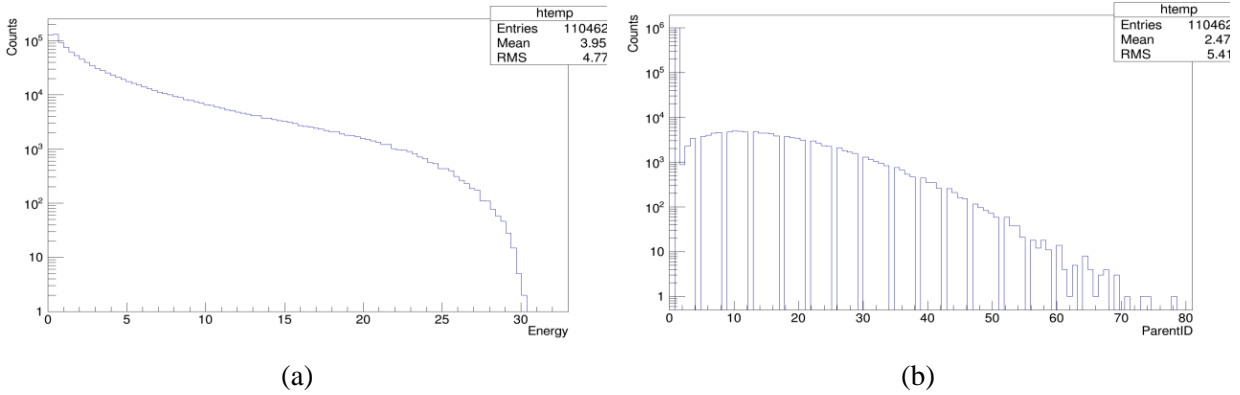

|  (a) | (b) |

**Figure 7.** (a) Energy spectra of bremsstrahlung photons simulated by Geant4 toolkit. (b) The distribution of the generations of the photons in the activation sample.

*Reaction rate density: tabulating photon flux with historical cross section*

Six product radioisotopes ($^{47}$Ca, $^{57}$Ni, $^{65}$Zn, $^{84}$Rb, $^{122}$Sb, $^{139}$Ce) and their corresponding photonuclear reactions are selected to validate the activity equation in pulse irradiation. They were chosen based on the following facts: (1) they have clear interference-free energy lines; (2) concentrations of target nuclides are certifieded; (3) they are products of dominant photonuclear reactions; and (4) their atomic number ranges from low to high in the period table.

**Table 1.** Selected photon nuclear reactions and their EXFOR records.

| Reaction | Exfor Record | Cross section in Exfor | Reference |
|---|---|---|---|
| $^{48}$Ca(γ,n)$^{47}$Ca | M0636007 | $\sigma(\gamma, n)$ | O'Keefe1987 |
| $^{58}$Ni(γ,n)$^{57}$Ni | L0034003 | $\sigma(\gamma, X)$ | Fultz1974 |
| $^{66}$Zn(γ,n)$^{65}$Zn | L0164002 | $\sigma(\gamma, n)$ | Coryachev1982 |
| $^{85}$Rb(γ,n)$^{84}$Rb | L0027002 | $\sigma(\gamma, X)$ | Lepretre1971 |
| $^{123}$Sb(γ,n)$^{122}$Sb | L0035033 | $\sigma(\gamma, n) + \sigma(\gamma, n+p)$ | Lepretre1974 |
| $^{140}$Ce(γ,n)$^{139}$Ce | M0367005 | $\sigma(\gamma, n)$ | Beljaev1991 |

Original data of cross section $\sigma$ is in exchange format (EXFOR), which contains an extensive compilation of experimental nuclear reaction data [14, 15]. Table 1 shows the EXFOR records of the selected photon nuclear reactions. Photon flux $\varphi$ is the product of electron flux $\varphi_e$ and phton yeild $Y$. $Y$ is obtained from photon shower simulations by the ratio of the number of photons in a certain energy bin in the target to total incident electrons. Number of photons in certain energy bins were counted by the histogram function in statistical package R [16]. Cross-section data and photon flux in different energy bins were tabulated, and their products are summed up to get the reaction rate density. Table 2 gives an example of the tabulate process for $^{140}$Ce(γ,n)$^{139}$Ce reaction.

In Table 2, $E$ is the energy in MeV, $\sigma$ is cross section in mb, $\Delta\sigma$ is uncertainty of cross section in mb, "Energy bins" is the energy range centered on $E$, $\varphi_e$ is number of electrons per second, $Y(E)$ is photon yield in the energy range per square centimeter, $\varphi(E)\sigma(E)$ is the reaction rate density in the energy range in $cm^{-2}s^{-1}$, $\Delta\varphi(E)\sigma(E)$ is the uncertainty of reaction rate density in the energy range, and $\int_{E_{thres}}^{E_{max}} Y(E)_p \sigma(E)_p dE$ is reaction rate density for the complete energy range.

**Table 2.** Reaction rate density and its uncertainty for $^{140}$Ce(γ,n)$^{139}$Ce reaction.

| $E$ | $\sigma$ | $\Delta\sigma$ | Energy bins | $Y(E)$ | $\varphi(E)\sigma(E)$ | $\Delta\varphi(E)\sigma(E)$ |
|---|---|---|---|---|---|---|
| 7.86 | 0.1 | 0.86 | 7.86-7.98 | 0.0001279 | 1.279E-32 | 1.1E-31 |
| 7.98 | 0.08 | 0.9 | 7.98-8.1 | 0.0001313 | 1.050E-32 | 1.18E-31 |
| 8.1 | 0.13 | 0.88 | 8.1-8.22 | 0.0001175 | 1.528E-32 | 1.03E-31 |
| 8.22 | 0.01 | 0.88 | 8.22-8.34 | 0.0001221 | 1.221E-33 | 1.07E-31 |
| 8.34 | 0 | 0.9 | 8.34-8.46 | 0.0001199 | 0 | 1.08E-31 |
| … | … | … | … | … | … | … |
| 21.3 | 55.78 | 7.14 | 21.3-21.42 | 0.0000165 | 9.2037E-31 | 1.1781E-31 |
| 21.42 | 54.7 | 7.27 | 21.42-21.54 | 0.000018 | 9.846E-31 | 1.3086E-31 |
| 21.54 | 53.75 | 7.76 | 21.54-21.66 | 0.0000191 | 1.02663E-30 | 1.48216E-31 |
| 21.66 | 52.89 | 8.85 | 21.66-30 | 0.0000474 | 2.50699E-30 | 4.1949E-31 |
| | | | Sum $\int_{E_{thres}}^{E_{max}}\varphi(E)\sigma(E)dE=$ | 6.619E-28 | ± | 1.99E-29 |

*Numerical simulation with MATLAB*

A live script in MATLAB was written to numerically imitate pulse superimposition in the irradiation [17]. All the parameters in scripts are originated from the real experiments conducted in previous experimental session. The codes of two loops below have been applied with the algorithms in the mathematical session. As we expected, the outputs of numerical simulations from two different algorithms are almost the same[1] and they agree with the calculation result from the new equation.

```
%% Derivation: numerical simulation with loops
% * Derivation 1: Superimposition each pulse at moment t_i
N1 = N_0*phi_p_sigma*(1-exp(-lambda*t_p))*exp(-lambda*(t_i-t_p))/lambda;
for i = (1:m-1) % from 1 to m, run loop, adding exp(-lambda*T) each time
    %disp(i)
    N1 = N1 + N_0*phi_p_sigma*(1-exp(-lambda*t_p))*exp(-lambda*(t_i-t_p-i*T))/lambda;
end
A_p_derivation1 = lambda*N1*exp(-lambda*t_r)  % exp(lambda*t_r) comes from the decay in the last pulse
% * Derivation 2: Addition of nuclides generated by each pulse one by one
N2=(N_0*phi_p_sigma/lambda)*(1-exp(-lambda*t_p))*exp(-lambda*(T-t_p));
for i = (2:m) % from 2 to m, run loop, adding exp(-lambda*T) each time
    %disp(i)
%N2 = N2 + (N2*exp(-lambda*t_p)+ (N_0*phi_p_sigma/lambda)*(1-exp(-lambda*t_p)))*exp(-lambda*(T-t_p))
    N2= N2 + (N_0*phi_p_sigma/lambda)*(1-exp(-lambda*t_p))*exp(lambda*t_p)*exp(-lambda*(i*T));
end
A_p_derivation2 = lambda*N2*exp(-lambda*t_r)
%%
```

## RESULTS AND DISCUSSIONS

*Have the experiments confirmed the validity of the general activity equation in pulse irradiation?*

Table 3 is the activity comparison between the theoretical predictions from the Equation (27) and the experimental values at the end of irradiation in the photon activation experiments. Experimental values are obtained from measured activities divided by the factor of decay $e^{-\lambda t_d}$. Discrepancy is the difference between experimental values and theoretical predictions in percentage. Z-score is the distance from the sample mean to the population mean in units of the standard error. One can see that theoretical activities are in the same order of magnitude as the measured activities. However, the discrepancy of predicted ranges between 20% and 40%. And all the predicted values are systematically larger than the real experimental values. This is understandable given by primarily two reasons: (1) the cross-section data applied are systemically higher. Because of the lacking cross section data of (g, n) reactions, some cross-section data employed is for $(\gamma, n)$ and $(\gamma, n) + (\gamma, n + p)$ (see Table 1); (2) simulated photon flux is usually higher than the real situation. Historical experiments and computer simulations with different Monte Carlo codes have shown that simulated flux is usually higher than the real experimental flux by around 20%, varied by different experimental setups and simulation programs [18]. Besides these two dominating causes, the discrepancy may also be contributed by a combination of several factors, such as beam emittance, uncertainty of beam current, beam loading, beam wandering, energy dissipation in experimental setup, efficiency measurements, etc.

---
[1] The slightly difference is originated from the inaccuracy of floating-point arithmetic.

**Table 3.** Activity comparison between theoretical prediction and experimental measurement.

| Reaction | Energy Line (KeV) | Half-life | Experimental Activity (Bq) | Theoretical Activity (Bq) | Discrepancy | Z-score |
|---|---|---|---|---|---|---|
| $^{48}Ca(\gamma,n)^{47}Ca$ | 1297.09 | 4d12h51m40s | 386±5 | 488±85 | 20.90% | 1.20 |
| $^{58}Ni(\gamma,n)^{57}Ni$ | 1377.63 | 1d11h36m40s | 238±16 | 316±39 | 24.68% | 2.00 |
| $^{66}Zn(\gamma,n)^{65}Zn$ | 1115.55 | 244d5h6m40s | 77±2 | 104±12 | 25.96% | 2.25 |
| $^{85}Rb(\gamma,n)^{84}Rb$ | 881.61 | 32d18h23m20s | 20±1 | 33±3 | 39.39% | 1.33 |
| $^{123}Sb(\gamma,n)^{122}Sb$ | 564.12 | 2d17h21m40s | 172±4 | 277±28 | 37.90% | 3.75 |
| $^{140}Ce(\gamma,n)^{139}Ce$ | 165.86 | 137d14h46m40s | 8.2±0.4 | 13.6±1.4 | 39.71% | 3.86 |

Figure 8a plots the values in Table 3. One can notice that experimental values are consistent with the predicted values despite some discrepancy. Figure 8b shows the statistical correlation between theoretical predictions and actual measurements. Z-test results have shown that the predicted values are statistically close to the experimental values. The correlation coefficient confirms that they are directly related (R ≈ 0.99289)[2]. Therefore, statistically, we are able to claim that the experiments of photon activation with LINAC has confirmed the validation of the new equation in pulse irradiation.

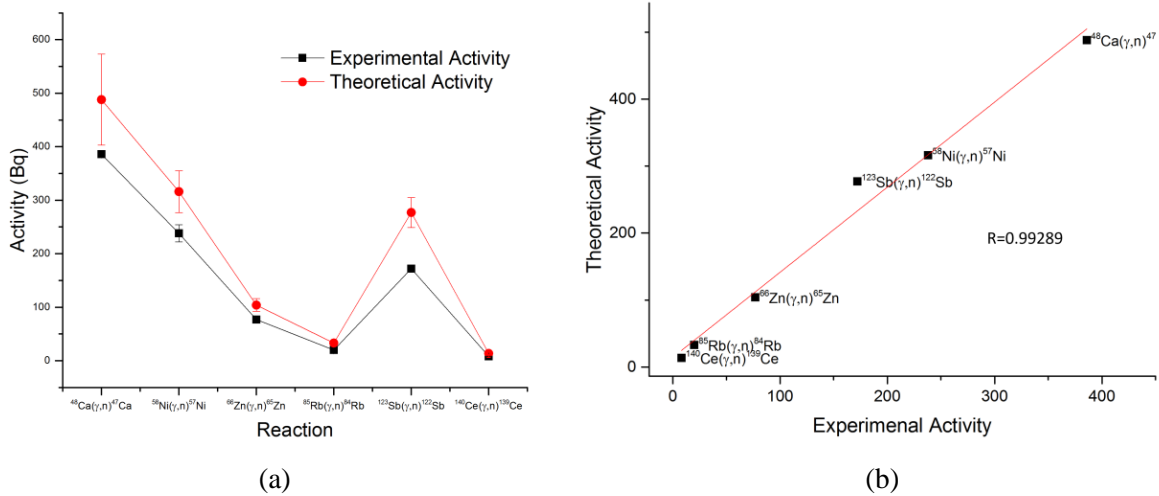

**Figure 8.** (a) Activity comparison between theoretical prediction and experimental measurement. (b) Z-test for activities from theoretical prediction and actual measurements.

*Comparison: pulse irradiation vs. continuous irradiation*

If the new equation was confirmed by the experiments, does it mean the traditional practice with continuous assumption are invalid and the foundation for photon activation analysis were built on sand? Fortunately, the answer to this question is no. The discrepancy of current and traditional equations is negligible in most photon activation cases. As mentioned before, the traditional method to calculate the activity is based on continuous assumption, which means $\varphi_p t_p = \bar{\varphi} T$ and

$$A_{continous} = N_0 \bar{\varphi} \sigma (1 - e^{-\lambda t_i}) = N_0 \varphi_p \sigma (1 - e^{-\lambda t_i}) \cdot \frac{t_p}{T} \tag{28}$$

---

[2] The linear fit parameters are: A(intercept) ≈ 14.42±15.28; B(slope) ≈1.271±0.076; Pearson's R ≈ 0.99289, R-Square(COD) ≈ 0.98583. In analysis of variance (ANOVA), F Value ≈ 278.29, (Prob>F) ≈7.565E-5.

Dividing (27) by (28), one gets the ratio $\zeta$ between the activities of pulse irradiation and continuous assumption is

$$\zeta = \frac{A_{pulse}}{A_{continuous}} = \frac{T}{t_p} \cdot \frac{1-e^{-\lambda t_p}}{1-e^{-\lambda T}} = \frac{T}{t_p} \cdot \frac{1-e^{-\frac{\ln 2}{\tau} t_p}}{1-e^{-\frac{\ln 2}{\tau} T}} \tag{29}$$

One can notice that $\zeta$ value has nothing to do with the total irradiation time $t_i$, but closely related to the ratio of $T/t_p$. Given by the experimental parameters (the pulses are very dense and the half-life of the product nuclides of interest are relatively long comparing with the whole irradiation time) in the traditional photon activation, the $\zeta$ value is almost equal to 1 without exception. There is no significant change either following the continuous assumption or the current equation of pulse irradiation[3].

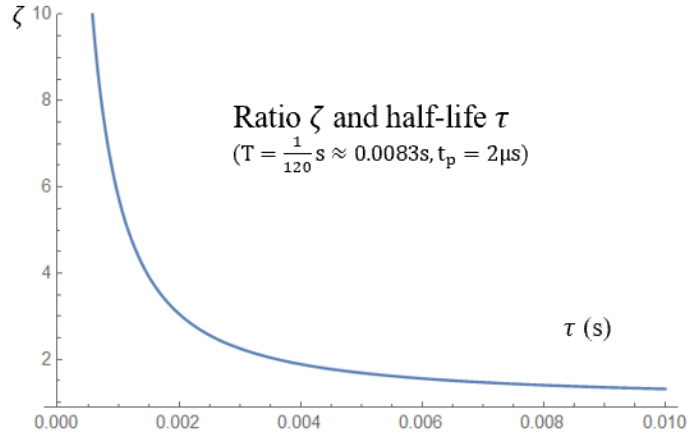

**Figure 9.** Activity comparison between pulse assumption and continuous assumption for radioisotopes with very short half-life.

Then, the question is: why bother to use the new equation? We prefer to use current equation because: (1) it is a generic equation for pulse irradiation. It clearly distinguishes the pulse up-time and down time and acknowledges the fact that the accelerator has its duty cycle. Logically and mathematically, the current equation is correct and avoids the unnecessary approximation of continuous irradiation. The discrepancy of the current and traditional equations may not be significant for photon activation, but it may be significant in some other cases, such as medical isotope production, radiation dose calculations, and nuclear physics. (2) In some extreme cases, the discrepancy of these two equations might be quite significant. For example, Figure 9 plots the relationship of $\tau$ and $\zeta$ in Mathematica within a typical photon pulse irradiation [19]. One can see that $\zeta$ changes its values significantly once half-life $\tau$ is less than the pulse period $T$: the shorter the half-life of radionuclide, the larger the ratio $\zeta$, with a maximum value of $\zeta$ close to the ratio of $T/t_p$. When the half-life and pulse period are in the same scale, the ratio is close to 1. Using traditional continuous assumption for those isotopes will not be accurate enough. However, this observation does not impact too much on nuclear activation analysis since what really counts in traditional radioanalytical practice is the ratio of activities, not the absolute activities of the same radioisotope in sample and reference [4, 20, 21].

---

[3] Philosophically speaking, the argument of continuous assumption or pulse irradiation is a new example of ancient Zeno's paradoxes.

*Further discussions and limitations of the new equation*

The relationship between pulse period $T$ and half-life $\tau$ of product nuclides not only plays a dominant role in the ratio $\zeta$, but also significantly impacts the final activity of the product nuclides. From the denominator $e^{\lambda T} - 1$ of Equation (25), one can notice that: if $T > \tau$, then the denominator $e^{\lambda T} - 1$ equals $e^{\frac{ln2}{\tau} \cdot T} - 1$, which is larger than 1, the whole fraction is smaller than $\lambda N_0 \phi \sigma e^{\lambda t_p}(1 - e^{-\lambda t_p})(1 - e^{-\lambda t_i})$. On the contrary, if $T < \tau$, the denominator is less than 1, and the result is larger than $N_0 \phi \sigma (1 - e^{-\lambda t_p}) e^{\lambda t_p}(1 - e^{-\lambda t_i})$. Thus, the turning point of activity is decided by the relationship of pulse period $T$ and the half-life $\tau$ of the product nuclides. If the pulse period $T$ is significantly shorter than the half-life $\tau$ of product nuclides, one can expect a substantial increase of final activity in pulse irradiation.

Although the current equation is a general equation for pulse irradiation, it can be easily broadened its usage to ion beams, reactors operating in pulse mode, and radiation dose calculation, it has its limitations as well: First of all, it is based on rectangular wave assumption of pulses. In some practical cases, sinusoidal wave assumption might be more accurate. Secondly, it is based on the assumption that the burn-up of target isotope can be negligible in irradiation. If the burn-up cannot be ignored, the current equation of pulse irradiation needs to make some adjustments according to the Bateman's Equations [22]. If one considers transient equilibrium, secular equilibrium, and other details in decay kinetics, the final activation equation will be more complicate than the current equation. However, the idea of applying geometry theories to mimic the activity superimposition is still valid.

## CONCLUSIONS

A novel activity equation for pulse irradiation was derived mathematically with the assistance of geometry series, and then it was confirmed numerically by MATLAB codes, and finally validated experimentally by photon activation conducted via a short-pulsed LINAC. The comparison between the new equation (based on pulse irradiation) and the traditional equation (based on an approximation of continuous irradiation) indicates that their discrepancy is negligible in most cases of photon activation, but it could be significant under certain conditions. The limitations of the new activity equation are discussed as well.

## ACKNOWLEDGMENTS

This research is supported by the U.S. Department of Energy, Office of Environmental Management (EM), MSIPP program under TOA # 0000272361. The author would like to thank Dr. Sadiq Setiniyaz of the Korea Atomic Energy Research Institute for his insightful comments.